\newcommand{\Msolar}{{\rm M_{\odot}}}   
\newcommand{\betacrit}{\beta_{\rm crit}}  

\documentclass{iaus}
\usepackage{graphicx}
\usepackage{subfigure}

\title[Resolution effects on fragmentation] 
{Non-convergence of the critical cooling timescale for fragmentation of self-gravitating discs}

\author[Farzana Meru and Matthew R. Bate]   
{Farzana Meru$^{1,2}$
\thanks{farzana@astro.ex.ac.uk}
 \and Matthew R. Bate$^1$}

\affiliation{$^1$School of Physics, University of Exeter, Stocker Road, Exeter, EX4 4QL, UK \\[\affilskip]
$^2$Institut f\"ur Astronomie und Astrophysik, Universit\"at T\"ubingen, Auf der Morgenstelle 10, 72076 T\"ubingen, Germany}

\pubyear{2010}
\volume{276}  
\pagerange{119--126}
\jname{The Astrophysics of Planetary Systems: Formation, Structure, and Dynamical Evolution}
\editors{A.C. Editor, B.D. Editor \& C.E. Editor, eds.}
\begin{document}

\maketitle

\begin{abstract}
We carry out a resolution study on the fragmentation boundary of self-gravitating discs. We perform three-dimensional Smoothed Particle Hydrodynamics (SPH) simulations of discs to determine whether the critical value of the cooling timescale in units of the orbital timescale, $\beta_{\rm crit}$, converges with increasing resolution.  Using particle numbers ranging from 31,250 to 16 million (the highest resolution simulations to date) we do not find convergence.  Instead, fragmentation occurs for longer cooling timescales as the resolution is increased.  These results certainly suggest that $\betacrit$ is larger than previously thought.  However, the absence of convergence also questions whether or not a critical value exists.  In light of these results, we caution against using cooling timescale or gravitational stress arguments to deduce whether gravitational instability may or may not have been the formation mechanism for observed planetary systems.
\keywords{accretion, accretion disks, gravitation, instabilities, hydrodynamics, methods: numerical, (stars:) planetary systems: formation, (stars:) planetary systems: protoplanetary disks}
\end{abstract}

\firstsection 
\section{Introduction}

There are two quantities that have historically been used to determine whether a self-gravitating disc is likely to fragment.  The first requires the stability parameter, $Q \lesssim 1$ (\cite[Toomre 1964]{Toomre_stability1964}), where $Q=\frac{c_{\rm s}\kappa_{\rm ep}}{\pi\Sigma G}$, $c_{\rm s}$ is the sound speed in the disc, $\kappa_{\rm ep}$ is the epicyclic frequency, which for Keplerian discs is $\approx \Omega$, the angular frequency, $\Sigma$ is the surface mass density and $G$ is the gravitational constant.  \cite{Gammie_betacool} showed the need for fast cooling and suggested that if the cooling timescale can be parameterised as $\beta = t_{\rm cool}\Omega = u \Big(\frac{{\rm d}u_{\rm cool}}{{\rm d}t}\Big)^{-1} \Omega$, where $u$ is the specific internal energy and ${\rm d}u_{\rm cool}/{\rm d}t$ is the total cooling rate, then fragmentation requires $\beta \lesssim \betacrit$, where the critical cooling timescale, $\betacrit \approx 3$ for a ratio of specific heats, $\gamma = 2$.  \cite{Rice_beta_condition} performed 3D SPH simulations to show the dependence of $\betacrit$ on $\gamma$: for discs with $\gamma = 5/3$ and $7/5$, $\beta_{\rm crit} \approx 6-7$ and $\approx 12-13$, respectively.  The above authors show that the cooling condition is equivalent to a maximum gravitational stress that a disc can support without fragmenting.  Recently, \cite{Meru_Bate_fragmentation} suggested that $\betacrit$ depends on the disc and star conditions.

We carry out a thorough convergence test of the value of $\beta_{\rm crit}$ required for fragmentation using a 3D SPH code.  We simulate a $0.1 \Msolar$ disc surrounding a $1 \Msolar$ star, spanning a radial range, $0.25 \le R \le 25$~au, using $\gamma = 5/3$.  The initial surface mass density and temperature profiles are $\Sigma \propto R^{-1}$ and $T \propto R^{-1/2}$, respectively, and initially, $Q \gtrsim 2$ everywhere.  We carry out simulations with this disc setup using 31,250, 250,000, 2 million and 16 million particles.  We simulate the discs using various values of $\beta$ to determine $\beta_{\rm crit}$ at different resolutions.

\section{Results}

\begin{figure}
  \begin{minipage}[b]{0.5\linewidth} 
    \centering
    \includegraphics[height=6.0cm]{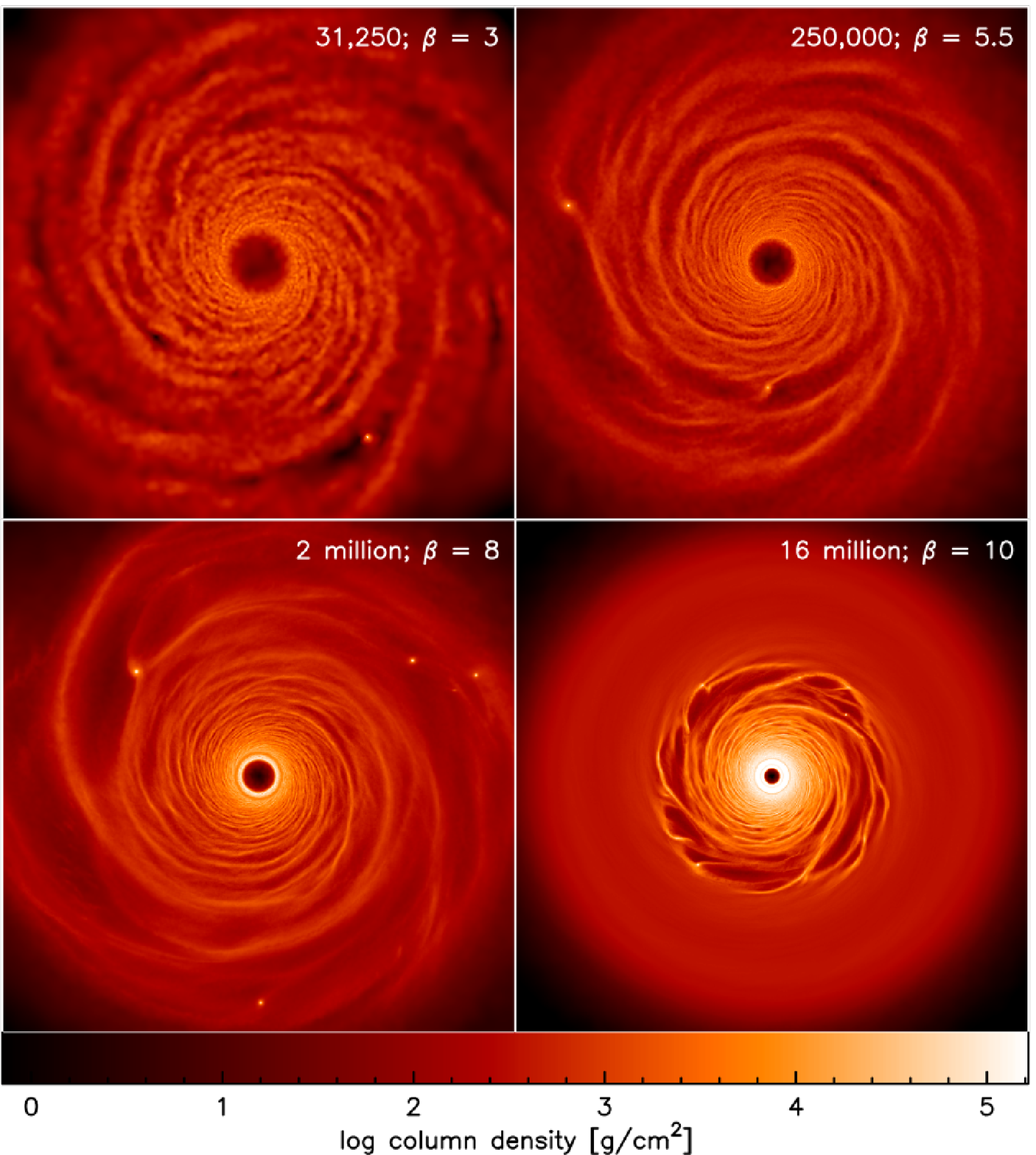}
  \end{minipage}
  \hspace{0.1cm} 
  \begin{minipage}[b]{0.5\linewidth}
    \centering
    \includegraphics[width=6.0cm]{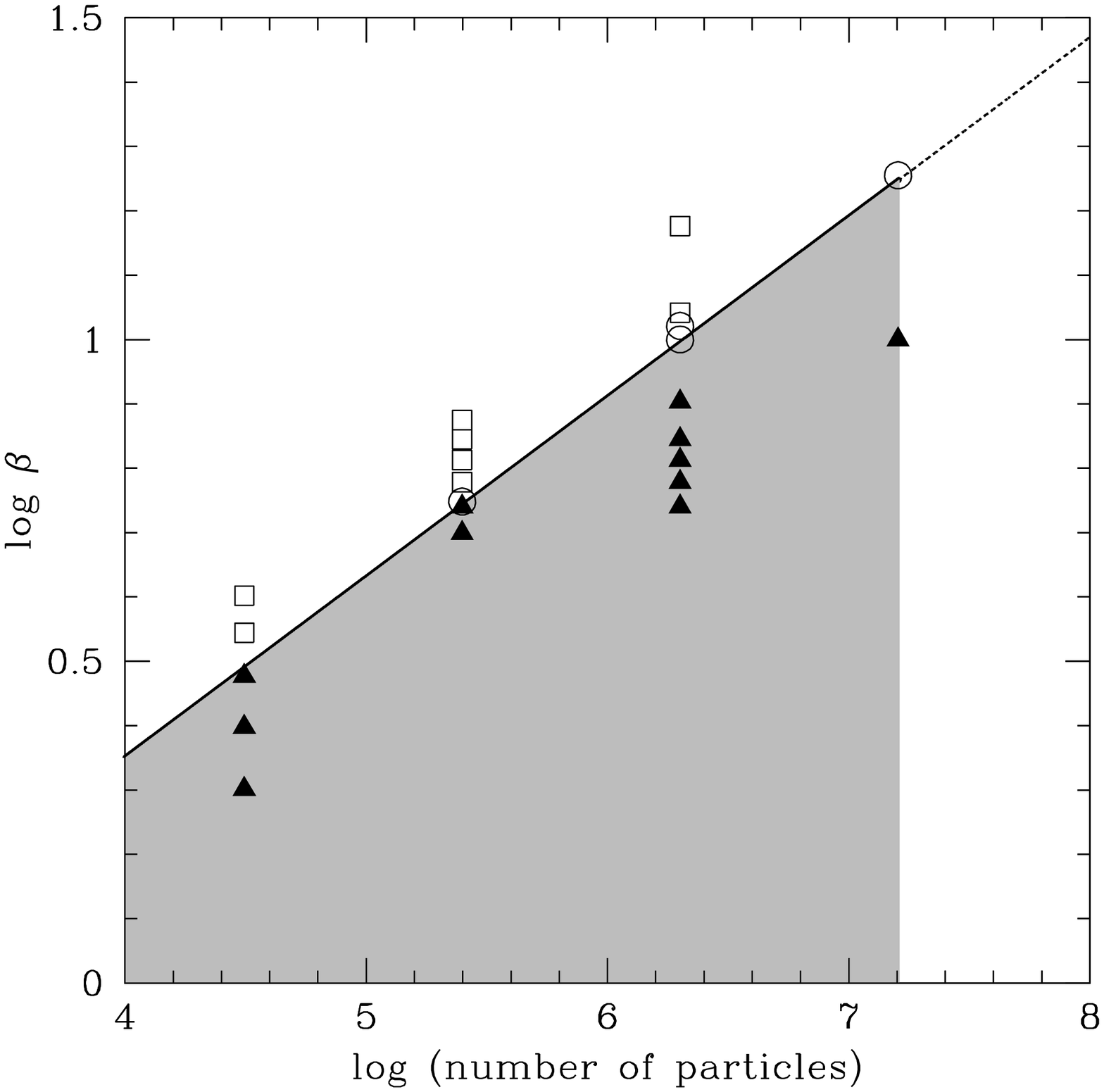}
  \end{minipage}
    \caption{\textbf{a.}~Surface mass density rendered images of the fragmenting discs with 31,250, 250,000, 2 million and 16 million particles.  At higher resolution, the disc can fragment for larger $\beta$.  The axes scale from -25~au to 25~au in both directions.  \textbf{b.}~Graph of $\beta$ against resolution of the non-fragmenting (open squares), fragmenting (solid triangles) and borderline (open circles) simulations (defined as discs that fragment but quickly shear apart with no further fragmentation).  The solid black line divides the fragmenting and non-fragmenting cases and the grey region is where fragmentation can occur.  The graph shows no evidence of convergence with resolution.  The thin dotted line shows how the trend will continue if convergence is not reached with even higher resolution.  If convergence can be achieved, the dotted line would follow a flatter profile.}
    \label{figure}
\end{figure}

Figure~\ref{figure}a shows images of the fragmenting discs at various resolutions.  We see that fragmentation occurs for higher values of $\beta$ as the resolution increases.  Figure~\ref{figure}b summarises all the simulations performed.  With the data that is available, the dividing line between the fragmenting and non fragmenting cases increases linearly with linear resolution and therefore \emph{convergence has not been reached}.

\section{Numerical and observational implications}

The lack of convergence certainly shows that the critical value of the cooling timescale is, at the very least, longer than previously thought.  However, it also opens up the possibility that there may be no value of $\beta$ for which such a disc can avoid fragmentation, given sufficient resolution.  If this is the case, it suggests that the problem may be ill posed.  In other words, it may not be possible for a disc to settle into an equilibrium where there is a balance between heating from gravitational instabilities and a {\it simple} imposed cooling timescale.  This implies that a self-gravitating disc that cools at a rate given by $\frac{{\rm d}u}{{\rm d}t}=\frac{u \Omega}{\beta}$, and is only heated by internal dissipation due to gravitational instabilities, may not be able to attain a self-regulated state and will always fragment, regardless of the value of $\beta$.  This re-opens the question of what the criterion for fragmentation of a self-gravitating disc really is and in addition, where in a disc fragmentation can realistically occur.  These results cast some serious doubts on previous conclusions concerning fragmentation of self-gravitating discs.  In addition, since cooling timescale arguments can be used to determine at what radii in a disc fragmentation can occur (e.g. Rafikov 2009; \cite[Clarke 2009]{Clarke2009_analytical}), the results presented here need to be considered when making conclusions as to whether observed planetary systems may or may not have formed by gravitational instability.

\end{document}